# High-contrast scattering microscopy in thick tissue with back-illumination interference tomography

GREGORY N. MCKAY,[1] JEROME MERTZ,[2] AND NICHOLAS J. DURR[3,*]

[1]*Department of Surgery, The Johns Hopkins University School of Medicine, Baltimore, MD, USA*
[2]*Department of Biomedical Engineering, Boston University, Boston, Massachusetts 02215, USA*
[3]*Department of Biomedical Engineering, Johns Hopkins University (JHU), 3400 N. Charles Street, Baltimore, MD 21218, USA*
[*]*ndurr@jhu.edu*

**Abstract:** Biological tissues are composed of discrete compartments of biochemical media that often exhibit subtle differences in refractive index. Light propagating through these compartments partially diffracts in a forward direction with a $\pi/2$ phase shift. We introduce a microscopy technique to image this scattering signal in thick tissues, called back-illumination interference tomography (BIT). An incoherent source is demagnified and imaged past the focal plane of a high numerical aperture objective lens, producing a small, semicoherent source of backscattered light. This backscattered light undergoes a phase inversion over the narrow depth of field of the microscope, providing interference contrast to weakly scattering objects at the focal plane. BIT offers a fundamentally different source of contrast to conventional illumination and oblique back-illumination microscopy. Compared to these techniques, we show that BIT improves contrast to blood cells *in vitro* in microfluidic chambers and *in vivo* in a human capillary. Finally, we apply BIT to unstained, unlabeled bulk human tissue *ex vivo* and compare side-by-side to adjacent frozen sections stained with Hematoxylin and Eosin. These results demonstrate the potential of BIT to provide high resolution, high speed, 3D imaging of unprocessed biological tissues.

## 1. Introduction

High-resolution imaging of unlabeled bulk tissues is critically important for characterizing biological tissues *in vivo*. A wide variety of imaging technologies have been introduced over the last few decades to address this need. Optical coherence tomography (OCT), which has gained widespread adoption in medical fields such as ophthalmology, provides depth-sectioning and label-free imaging through low-coherence interferometry [1]. Subsequently, full field optical coherence tomography (FF-OCT) was developed to eliminate the need for raster scanning [2,3]. Scattering-based confocal microscopy, in which out-of-focus scattered light is rejected to achieve optical sectioning, has shown promise in dermatology, and adherent leukocytes have even been observed in human vasculature with this technique [4–8].

Since Zernike's initial observation and publication on phase contrast, phase contrast microscopy techniques have become ubiquitous in biological research due to their label-free nature, negligible phototoxicity, high speed, and sensitivity to weakly scattering objects [9,10]. Quantitative phase imaging (QPI) enables the measurement of the refractive index of unaltered tissue structures [11]. Transmission-based illumination with thin, weakly scattering samples has long dominated the field of phase contrast microscopy to avoid the influence of multiple scattering that disrupts phase. Techniques such as gradient light interference microscopy (GLIM) and epi-GLIM have sought to address this using white light interferometry, a Nomarski approach, and phase shifting. While this technique enables quantitative phase imaging in bulk tissue, its speed is limited to around sixteen frames per second due to its liquid crystal variable retarder [12,13].

When introduced in 2012, oblique back-illumination microscopy (OBM) provided a simple and fast approach to generate phase contrast images in bulk turbid media [14]. OBM uses laterally offset illumination sources to produce backscattered light that passes through the focal plane at a net oblique angle, encoding lateral phase gradients at this plane to intensity changes at the detector. Since this seminal work, the field has been further developed to enable quantitative phase imaging using deconvolution [15]. While imaging bulk pathology and stationary *in vivo* samples has shown promise with qOBM [16], the requirement of multiple triggered sources and computational expense of reconstruction hinders the speed of this technology, making it difficult to apply to *in vivo* blood cell imaging [17].

In this article, we introduce a new epi-mode microscopy technique called back-illumination tomography (BIT). BIT generates contrast to weakly scattering objects using back-scattering of a spatially confined source imaged on-axis, beyond the objective focal plane. The technique operates similar to optical transmission tomography (OTT) demonstrated recently and explained by the Gouy phase shift [18]. We offer an alternative description of the contrast mechanism via the expected amplitude point spread function for a partially coherent system described by Streibl [19]. BIT's epi-mode geometry enables its use in bulk, turbid media, and its simplicity and speed make it appealing for *in vivo* blood cell imaging and bulk tissue histopathology. A high numerical aperture (NA) objective allows both the creation of spatially coherent backscattered light and a short axial imaging point spread function, which, in combination, allows the visualization of fine tissue structures with optical sectioning. We first demonstrate agreement between simulation based on Streibl's model and experimental data from imaging scattering $TiO_2$ particles. Next, we image human blood cells *in vitro* in microfluidic chambers to characterize the relationship between the illumination source image depth and BIT contrast. Finally, we demonstrate two applications of BIT: (1) imaging flowing human blood cells with high speed and high resolution *in vivo*, and (2) imaging unprocessed bulk human tissue specimens.

## 2. Methods

### 2.1. Optical System

The BIT optical system is outlined in Fig. 1. It consists of an LED placed within the back focal length (BFL) of a condensing lens (CL, Thorlabs ACL25416U), a distance $d$ away from the lens vertex. Note that the distance the LED is placed from the BFL of CL is $\delta z = f_{CL} - d$. The light source is mounted on an XYZ translation stage by affixing the LED heatsink to a Thorlabs kinematic mount using J-B Weld SteelStik. Diverging light is relayed through the condensing lens through a 50:50 non-polarizing beamsplitter (BS, Thorlabs CCM1-BS013), undergoing a reverse pass through an infinity corrected microscope objective (obj., Nikon 40X 1.15 NA APO LWD WI $\lambda$S). Note that purple dashed lines in Fig. 1 represent conjugate planes, from which the LED is axially displaced. This configuration creates a demagnified image of the LED a distance $\delta z'$ beneath the focal plane of the microscope objective. Back-scattered light occurring from the high intensity region defined by the LED image creates net on-axis, back-illumination from a spatially confined source. The coherence factor ($\gamma$) is determined by Eq. 1, where $NA = n * sin(\theta)$, for the relatively high objective collecting angle ($\theta_{obj.}$) and the relative low source illumination angle ($\theta_{illum.}$).

$$\gamma = NA_{illum.}/NA_{obj.} \tag{1}$$

Using a thin lens approximation with $\delta z = 8$ mm, we estimate $\delta z' = 200$ $\mu$m, $|D_{LED}| = 100$ $\mu$m, and the resulting coherence factor is $\gamma = 0.25$ for green LED illumination (Supplementary Figures S1 and S2). Light is collected by the high NA objective, reflected off the beamsplitter, and imaged via a tube lens (TL) onto a CMOS sensor (pco.edge sCMOS 5.5 or The Imaging Source DMK 33UX252, for the case of *in vivo* blood cell imaging Fig. 4). For experiments that require axial scanning, the Nikon objective was mounted on a piezo collar (Thorlabs PFM450E) or XYZ translation stage (Thorlabs Nanomax MAX313D). Three different illumination source types were used in this study, including a 530 nm green LED (Luxeon Star Green Saber Z1 10mm, 92 lm, 500 mA), a 660 nm red LED (Luxeon Star Deep Red Saber Z1 10 mm, 325 mW, 500 mA), and a 530 nm green fiber source (Thorlabs M530F1 coupled to FT400UMT 0.39 NA, $\phi = 400$ $\mu$m). Note that data collected and presented in Fig. 2 and Fig. 4 utilized the 530 nm green LED, Fig. 3 utilized the 530 nm green fiber source, and data collected and presented in Fig. 5 utilized the 660 nm LED source. The choice of source type was dictated by whether or not absorption information from hemoglobin was desired, steric hindrance that was eased by utilization of a fiber source allowing $\delta z = f_{CL}$ ($d = 0$ mm), and empirically observed optimization of contrast for pathologic tissue specimen in the red portion of the visible spectrum.

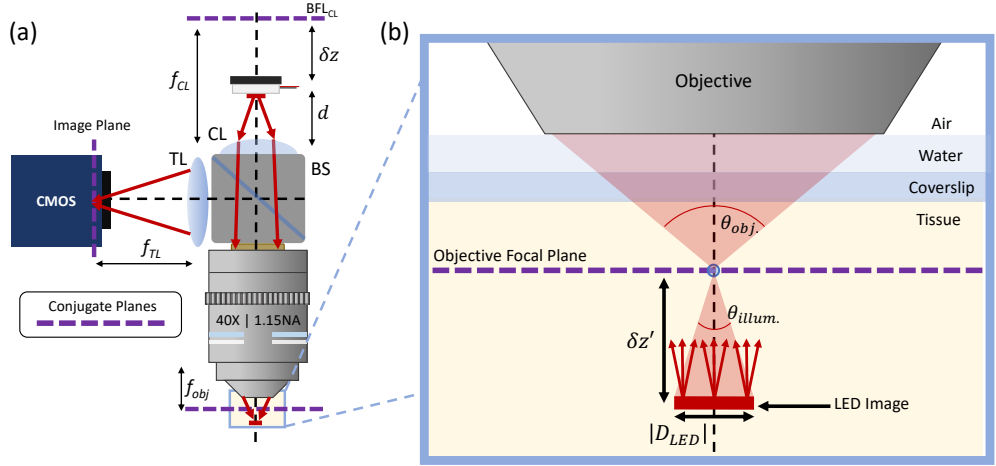

Fig. 1. (a) Back-illumination interference tomography (BIT) optical system. An LED placed within the focal length of a collecting lens ($f_{CL}$) produces diverging illumination through a beamsplitter (BS) that creates an LED image below the objective focal plane. Backscattered light is collected by the objective and imaged onto a CMOS via a tube lens (TL). (b) Enlarged region of interest highlighting how BIT achieves partially spatially coherent illumination through the demagnified LED image, located $\delta z'$ beneath the object plane. Backscattered light creates transmission-like illumination. Partially spatially coherent illumination is achieved due to $\theta_{obj.} > \theta_{illum.}$ ($\gamma < 1$).

### 2.2. BIT contrast mechanism

The 3D Optical Transfer Function (OTF) for a partially spatially coherent telecentric optical imaging system for weakly scattering objects was derived previously by Streibl and expanded upon by Sheppard & Mao [19, 20]. Due to the partially coherent, transmission-like illumination generated with BIT, this theory describes the contrast observed here well. In this model, a 3D weakly scattering object $t(\mathbf{v}_0)$ can be represented by its unscattered $\delta(u_0)$ and scattered radiation $t_1(\mathbf{v}_0)$,

$$t(\mathbf{v}_0) = \delta(u_0) + t_1(\mathbf{v}_0) \qquad (2)$$

where $\mathbf{v}_0$ represents the object-space vector comprised of optical coordinates $v_0$, $w_0$, and $u_0$, such that $v_0 = kx_0 sin(\theta_{obj.})$, $w_0 = ky_0 sin(\theta_{obj.})$, $u_0 = 4kz_0 sin^2(\theta_{obj.}/2)$. The object can be decomposed into its object spectrum $T_1(\mathbf{m})$,

$$t_1(\mathbf{v}_0) = \int T_1(\mathbf{m}) \exp(j\mathbf{m} \cdot \mathbf{v}_0) \, d\mathbf{m}, \qquad (3)$$

Where **m** is the Fourier-space angular frequencies corresponding to $\mathbf{v_0}$. Further, the real part of the object spectrum represents the absorptive variations and the imaginary part of the object spectrum represent phase variation

$$T_1(\mathbf{m}) = T_{1r}(\mathbf{m}) + iT_{1i}(\mathbf{m}). \tag{4}$$

The image intensity observed, $I(\mathbf{v})$, where **v** is the image-space optical coordinates conjugate to $\mathbf{v_0}$, is dependent on the object spectrum, the absorptive transfer function $C_A(\mathbf{m};\mathbf{0})$, and phase transfer function $C_P(\mathbf{m};\mathbf{0})$ as shown below.

$$I(\mathbf{v}) = C_D(\mathbf{0};\mathbf{0}) + \iint [C_A(\mathbf{m};\mathbf{0})T_r(\mathbf{m}) + C_P(\mathbf{m};\mathbf{0})T_i(\mathbf{m})\exp(j\mathbf{m}\cdot\mathbf{v})]\,d\mathbf{m}, \tag{5}$$

$$C_A(\mathbf{m};\mathbf{0}) = C_w(\mathbf{m};\mathbf{0}) + C_w^*(-\mathbf{m};\mathbf{0}) \tag{6}$$

$$C_P(\mathbf{m};\mathbf{0}) = C_w(\mathbf{m};\mathbf{0}) - C_w^*(-\mathbf{m};\mathbf{0}) \tag{7}$$

$$C_w(\mathbf{m};\mathbf{0}) = \frac{2}{\rho}\operatorname{Re}\left\{\left[\frac{1}{2}(1+\gamma^2) - \frac{\rho^2}{4} - \frac{r^2}{\rho^2} - \left|r - \frac{1}{2}(1-\gamma^2)\right|\right]^{1/2}\right\}, \tag{8}$$

Where $C_w(\mathbf{m};\mathbf{0})$ is determined by the generalized 6D partially coherent transfer function [20], $\rho = (m^2+n^2)^{1/2}$ represents transverse spatial frequencies, $r$ represents the axial spatial frequencies, and $\gamma$ is the coherence factor outlined by Equation (1). Importantly, for $\gamma < 1$, $C_P(\mathbf{m};\mathbf{0}) \neq 0$, meaning relative phase is observable, however for $\gamma \geq 1$, $C_P(\mathbf{m};\mathbf{0}) = 0$, becoming independent of $\gamma$, and only absorptive information is imaged. In addition, $C_P(\mathbf{m};\mathbf{0})$ is asymmetric about $r = 0$, ultimately leading to an intensity inversion about the objective focal plane. Though this effect is commonly observed in transmission microscopy, BIT is novel in that it enables epi-mode imaging of scattering objects using a simple illumination modification.

### 2.3. Scattering particle measurement and simulation

To evaluate the contrast generated in BIT, a simple scattering-only phantom was fabricated by mixing $TiO_2$ particles in a polydimethylsiloxane (PDMS) phantom. $TiO_2$ particles were dissolved at a concentration of 1 mg/mL to mimic tissue scattering of approximately $\mu'_s \sim 1.2\,\text{mm}^{-1}$ at 530 nm. With the 530 nm LED illumination placed at approximately $\delta z = 8$ mm, an axial scan of $TiO_2$ particles was acquired in 500 nm axial increments. The results of this experiment are shown in Fig. 2(a)-(b). Note that images are processed with a flat-field correction algorithm adopted from Ford et al., whereby images are divided by an 80-pixel Gaussian blurred version of themselves to generate a 32-bit corrected image, and contrast adjusted from [0.9,1.1] to improve visualization and generate near-uniform illumination across the FOV [14].

Following the model from [21], we used Eqs. 5 - 8 to simulate the axial intensity distribution for a 590nm spherical $TiO_2$ particle in PDMS with $\gamma = 0.25$ (Fig. 2(c)). The profile of this simulation is compared to experimentally measured imaging data in Fig. 2(d). Note that each of the images, experimental and simulated data, underwent intensity normalization using the flatfield corrected algorithm described above.

### 2.4. In-vitro static blood cell imaging experiment

Tissue-mimicking microfluidic chambers were fabricated following a previously developed protocol [22, 23]. Briefly, using spin-coating, SU8-3010 photoresist (PR) is deposited uniformly on a 76.2 mm silicon wafer (University Wafers #447) and exposed to UV light through a photomask in the shape of branching microvasculature (Fig. 3(a)). The PR is cured, producing

25 µm tall channels on the device mold. Separately, PDMS is doped with $TiO_2$ and India ink to approximate tissue optical properties ($\mu'_s = 1.7\,\text{mm}^{-1}$ and $\mu_a = 0.017\,\text{mm}^{-1}$ at 650 nm). The turbid PDMS is cast over the device mold, cured, cut out, and channels are cut through the device base with a blunt-tipped 20 gauge needle. Separately, a thin membrane of PDMS approximately 100 µm thick doped with $TiO_2$ and India ink at the same concentration as the device base, is spin coated on an additional silicon wafer coated with a soft-baked layer of PR (S-1813). The device base is plasma bonded to the thin membrane and the second silicon wafer is released from the finished device using an acetone bath to dissolve the soft-baked PR layer. This produces a PDMS microfluidic device with optical properties mimicking human tissue and 25 µm tall capillary channels embedded beneath 100 µm of turbid media.

Whole blood, purchased from ZenBio (SER-WB10ML-SDS), was loaded into a 1 mL syringe and connected to the microfluidic device using PEEK tubing and Luer adapters (IDEX 1569L, P-659, F-247, F-333NX). After initial blood flow was produced with a syringe pump, the flow was stopped and whole blood inside the microfluidic device was imaged while mostly stationary (only undergoing Brownian motion). For a fixed field of view containing a red blood cell (RBC) of interest, the distance from the 530 nm green fiber tip to the condensing lens was varied with an image taken at 2 mm increments of axial translation (Fig. 3(b)). The LED position was varied across the full axial range of interest, from lens vertex ($\delta z = f_{CL}$) to the back-focal plane of CL (critical imaging, $\delta z = 0$ mm). Plot profiles at the two extremes are shown. The experiment was repeated again, with a pair of white blood cells (Fig. 3(c)). Finally, for the white blood cells, with the fiber tip at critical illumination ($\delta z = 0$ mm), the fiber was laterally translated to generate comparative oblique-back illumination microscopy images [14, 17].

### 2.5. In vivo human blood cell imaging

The BIT microscope was inverted and mounted on an ophthalmic slit lamp housing to image the ventral surface of a human tongue to image blood cells flowing through superficial capillaries *in vivo*. Using a custom pneumatic suction objective cap published previously, the tongue was stabilized during imaging [17, 24]. Compared to previous studies, to reduce the bulkiness of the system, a smaller and lower-cost image sensor was used (The Imaging Source DMK 33UX252). Videos were acquired at 200 fps, with an exposure time of 1 ms, a gain of 19.7, and a gamma correction of 4. 530 nm green LED illumination was used to enable simultaneous acquisition of absorption and phase signal from blood cells. Human imaging was conducted with approval by the Johns Hopkins University Institutional Review Board (IRB00204985). Consenting patients placed their tongue on the objective cap for up to 30 minutes to complete an imaging session.

### 2.6. Ex vivo human pathologic tissue imaging

The BIT microscope was adapted for *ex vivo* tissue pathology imaging. A piezoelectric stage (Thorlabs PFM450E) was installed on the microscope objective to enable axial scanning. A matte black spray paint was applied to the internal brass collar of this stage to reduce stray reflected light. For comparative transmission images of sliced tissues, we added a Kohler illumination system with a white light LED (Luezon Star 6500K Saber Z1 10mm 154 lm @ 500 mA). A 660 nm LED was used for BIT imaging in epi-mode, as shown in Fig. 1. A flip mirror at 45° was installed after the beamsplitter and tube lens along with an RGB sensor (The Imaging Source DFK 38UX304) to enable color acquisition of Hematoxylin and Eosin (H&E) stained tissue samples. This approach enabled conventional microscope slides to be imaged with white illumination in transmission mode, and easily swapped for bulk tissue imaging with epi-illumination in BIT.

Discarded tissue specimens were obtained from Whipple procedures with approval by the Johns Hopkins University Institutional Review Board (IRB00204985). Tissue was frozen in O.C.T. compound (Fisher Healthcare Tissue-Plus O.C.T. Compound 23-730-571), sectioned in a cryostat with 15 µm slices, mounted on (+) charged 75 mm glass slides (Ted Pella Diamond

260382-1), stained following conventional H&E protocol (Abcam ab245880), and preserved using mounting media (Ted Pella Histomount 19479) and a coverslip (Ted Pella 260140). The remaining bulk tissue on the chuck was thawed in room temperature PBS, placed within a 35 mm glass bottom culture dish (Cellvis D35-28-1.5-N) and gently compressed from above with a coverglass window. This tissue is unstained and unsectioned bulk tissue, but the face imaged with BIT is directly adjacent to the tissue face that was sliced and stained for H&E imaging.

## 3. Results and Discussion

### 3.1. Scattering Particle Measurement and Simulation

To characterize the observed contrast generated by the BIT system, a simple scattering phantom consisting of $TiO_2$ particles in PDMS was imaged. Fig. 2 shows the results of this experiment. We observe that the scattering-only $TiO_2$ particles exhibit a bright-to-dark intensity variation with respect to the background intensity (Fig. 2(a)). To further highlight this, Fig. 2(b) shows one such $TiO_2$ particle with a ~595 nm measured diameter imaged at five distinct axial locations centered about the objective focal plane. When placed within the objective focal plane (-z defocus), the particle appears bright, as it is translated to the focal plane the contrast decreases, and when translated to beyond the focal plane (+z defocus), the same particle becomes dark. This behavior is expected for partially spatially coherent illumination of a scattering object.

Next, we simulated a 595 nm $TiO_2$ particle in PDMS ($\Delta n = 1.3$) with a coherence factor of $\gamma = 0.25$. Fig. 2(c) shows the resulting simulation in an axial xz plot. Taking axial plot profiles and mean intensity measurements through the axial focus of the measured particle (blue solid line) and the simulated particle (orange dashed line), we observe good agreement for the predicted axial range over which a particle of this size and index of refraction mismatch undergoes intensity inversion (Fig. 2(d)).

Discrepancies between the measured and simulated data are likely due to vibrations of the optical table dithering the axial and lateral position of the particle, multiple scattering events not accounted for in the simulation (which utilizes the first Born approximation), and that the $TiO_2$ particle imaged is not a perfect sphere (as is assumed in the simulation). Still, the agreement between the experiment and the simulation suggests that the signal arises from transmission-mode illumination with a partially spatially coherent source. The defocused, demagnified LED image beneath the particle produces transmission-like illumination from a spatially confined source due to net back-illumination.

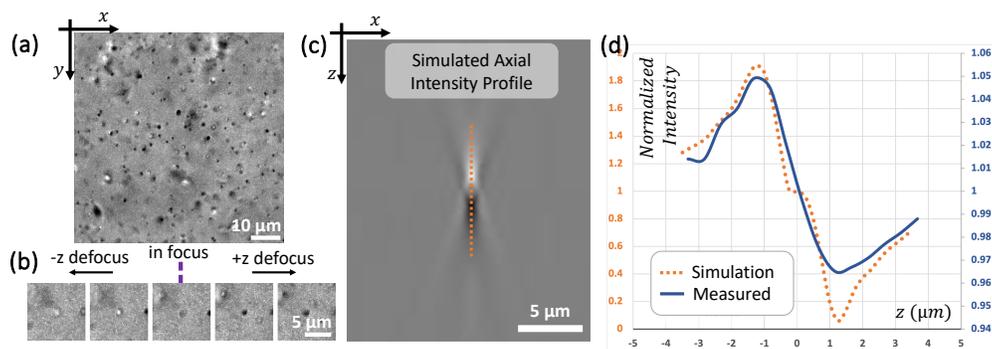

Fig. 2. (a) Scattering $TiO_2$ beads in polydimethylsiloxane (PDMS) phantom exhibit axially-dependent bright-dark contrast with BIT due to partially spatially coherent illumination. (b) Lateral images of a 595 nm $TiO_2$ particle at 500 nm axial increments about the objective focal plane demonstrate a transition from relative bright to dark intensity. (c) Simulated axial intensity profile for a 595 nm $TiO_2$ bead in PDMS with partially coherent illumination ($\gamma = 0.25$). (d) Normalized intensity vs. axial position for simulated particle (orange dashed, from (c)) and measured $TiO_2$ particle (blue solid, from (b)).

### 3.2. Imaging static blood cells

Blood cells were imaged *in vitro* microfluidic chamber emulating the geometry and scattering optical properties of a capillary in skin. With stationary bloodflow, the configuration of the BIT system was modified by adjusting the variable $\delta z$, which directly affects the coherence factor $\gamma$. Fig. 3(a) shows a schematic of the microfluidic chamber experiments utilized for this experiment. A field of view with a flat RBC showing its characteristic bi-lobed shape was selected for imaging (Fig. 3(b)). We observed the contrast of the blood cells to vary significantly, dropping off when $\delta z \lesssim 6$ mm, correspondingly when $\gamma \gtrsim 0.1$. Plot profiles across the same RBC full diameter are taken at the two extremes (green: $\delta z = f_{CL}$, red: $\delta z = 0$ mm, showing how contrast of the RBC membrane is enhanced with BIT illumination. At critical illumination ($\delta z = 0$ mm), it appears that the contrast is primarily from absorption of the larger volume of hemoglobin sequestered in the side lobes of the biconcave red blood cell.

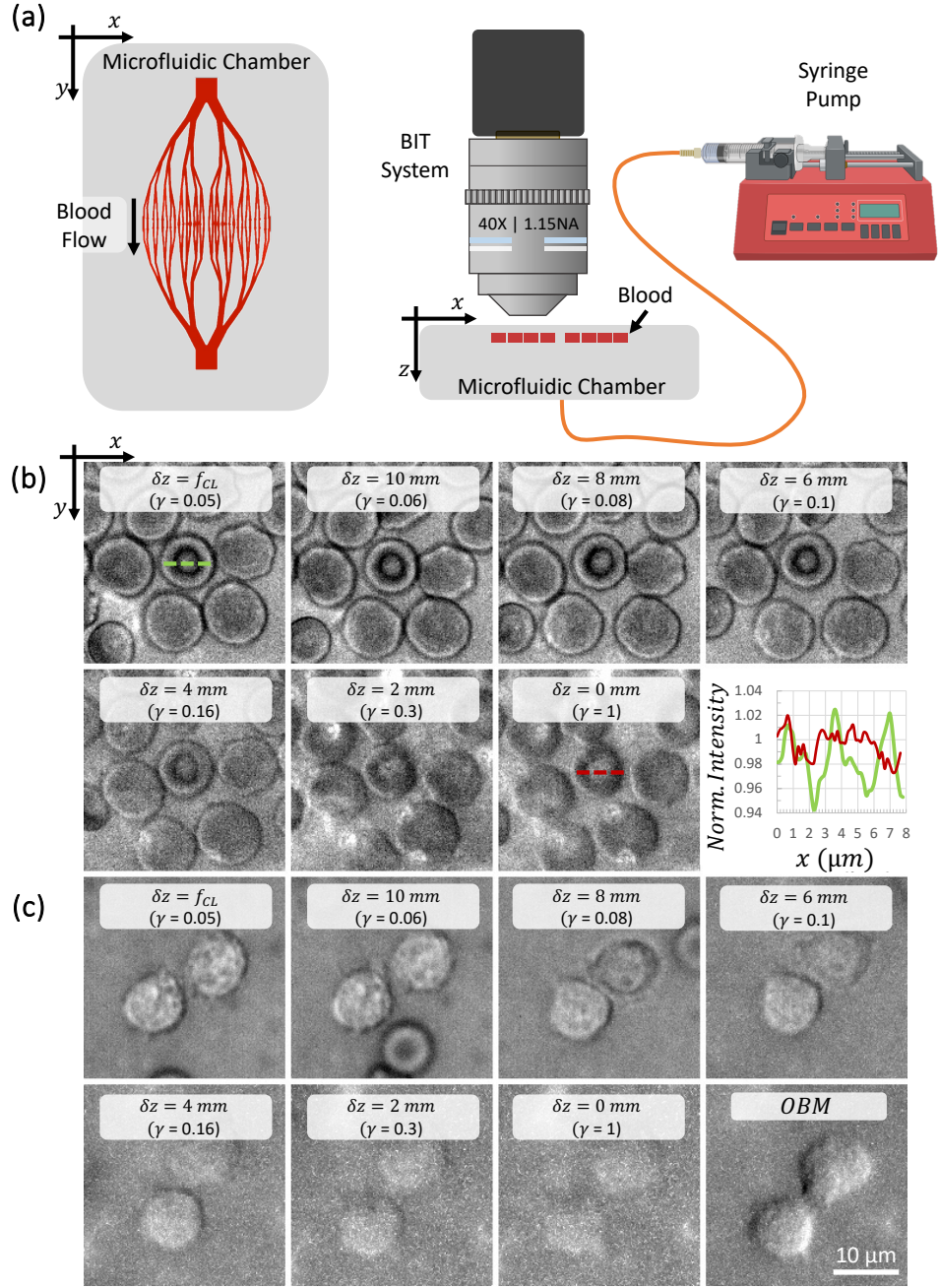

Fig. 3. (a) Schematic of microfluidic setup for *in-vitro* human blood cell imaging in a turbid medium. Flow was paused to enable imaging the same cells with varying $\gamma$. (a) A red blood cell (RBC) as the LED source is moved from the collecting lens vertex ($\delta z = f_{CL}$) to the collecting lens back focal plane (critical illumination, $\delta z = 0$ mm). The contrast of RBC membrane decreases as the LED moves closer to critical illumination and source coherence decreases (increasing $\theta_{illum.}$). This is highlighted with normalized intensity plot profiles taken across the RBC at $\delta z = f_{CL}$ mm (green), and $\delta z = 0$ mm (red). (c) Images of the same two white blood cells (WBCs) as the LED source is moved from the collecting lens vertex ($\delta z = f_{CL}$) to the collecting lens back focal plane (critical illumination, $\delta z = 0$ mm). At the critical illumination distance ($\delta z = 0$ mm), the LED was also translated laterally to generate oblique back-illumination microscopy (OBM) contrast.

### 3.3. In vivo human blood cell imaging

We investigated the application of BIT to in-vivo blood analysis by imaging capillaries in the ventral tongue of a human participant. Data from this experiment are shown in Fig. 4, with sequential frames from a video acquired at 200 frames per second. The video data from these experiments are included (visualization 1). Fig. 4(a) shows RBCs flowing through a capillary loop with the same RBC denoted with the red arrow. In Fig. 4(b), a smaller ROI is selected that tracks with the same RBC centered in the frame. The capillary vessel projects towards and then away from the tissue surface, perpendicular to the imaging plane. As the RBC passes through this capillary loop, its axial position with respect to the focal plane of the objective varies. Even though it is a primarily absorptive object at the 530 nm wavelength used in this experiment, the measured intensity of the RBC varies from bright to dark, just as observed with the $TiO_2$ particle experiments. The other major cellular components of blood are platelets and WBCs. Fig. 4(c)-(d) shows these components. A WBC passes through the capillary loop (white arrow), preceded by a platelet (blue arrow) and a plasma gap (visualization 2). BIT clearly resolves capillary RBCs, WBCs, and platelets with high contrast at high speeds, highlighting the potential of the technique for non-invasive, label-free blood analysis.

### 3.4. Ex vivo human pathologic tissue imaging

Lastly, we demonstrate the application of BIT in pathology, where label-free bulk tissue imaging could improve histopathology processing workflows and enable *in vivo* intraoperative assessment. We imaged adjacent regions of tissue specimens with conventional frozen section and H&E staining and label-free bulk tissue BIT imaging. The results of this experiment are shown in Fig. 4, with pancreas (a)-(b) and duodenum (c)-(d) tissues. Tissue types are paired between BIT and H&E column-wise, with zoomed in ROIs shown as highlighted green and red boxes in the right-most columns. BIT resolves nuclei and surrounding tissue structures, despite having no absorption-based contrast mechanism and without transmission illumination typically required for phase contrast techniques. Further studies across different tissue types, with both benign and diseased tissue, *in vivo* and *ex vivo* will further elucidate the potential of BIT in this field. With sufficient training data generated via this adjacent tissue imaging technique, an unsupervised generative artificial intelligence model could be trained to convert BIT images to virtual H&E stained images, as has been shown in OBM [16].

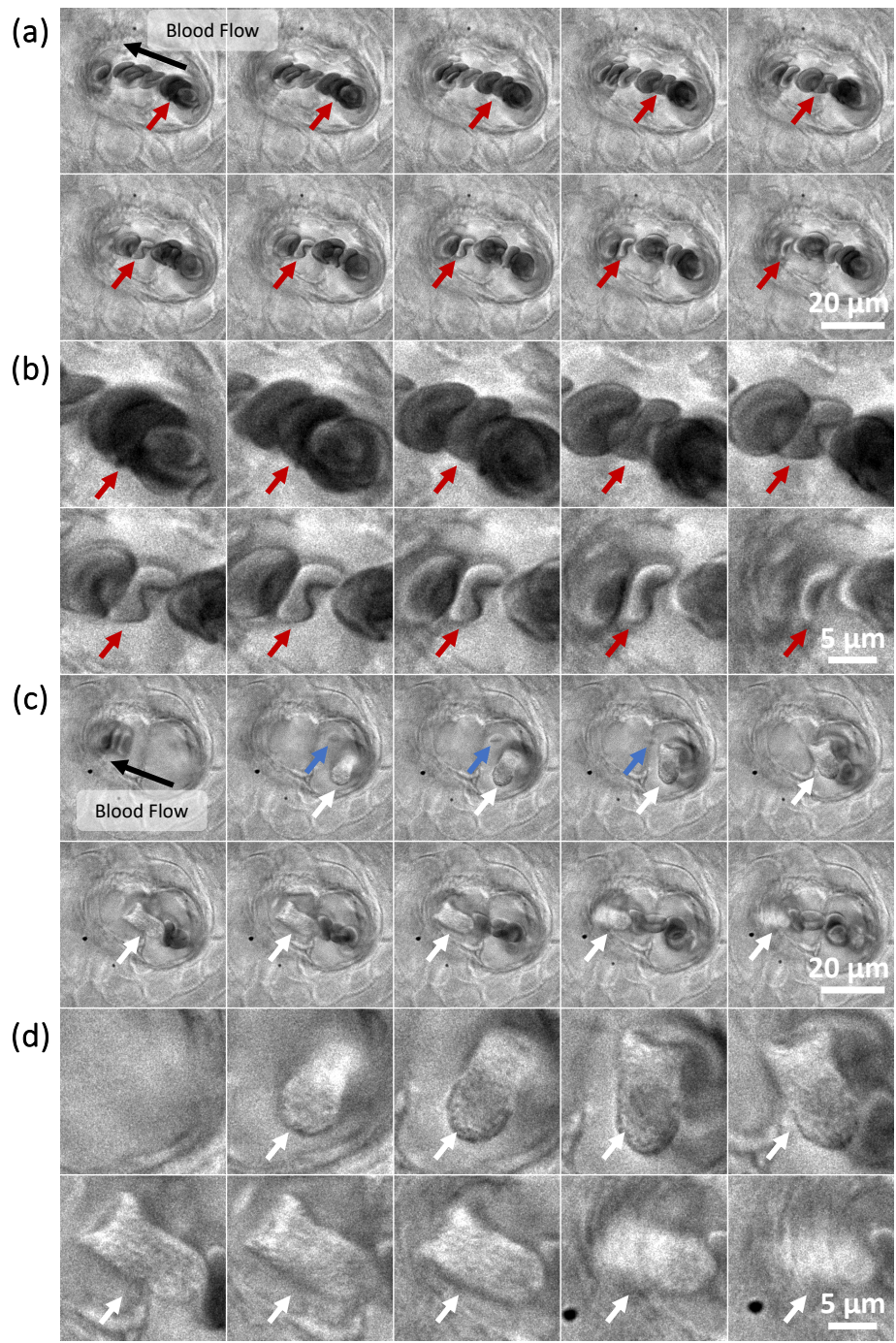

Fig. 4. *In vivo* imaging of human ventral tongue vasculature with BIT. (a) RBCs flowing through human capillary with red arrows denoting the same RBC tracked over time (Visualization 1). (b) Zoomed region of interest tracking the same RBC from (a) centered within the FOV. (c) WBC (white arrows) with preceding plasma gap and platelet (blue arrows) (Visualization 2). (d) Zoomed region of interest tracking the same WBC from (c) centered within the FOV. Each subfigure shows sequential frames from a video at 5 ms time increments, increasing from left-to-right then top-to-bottom.

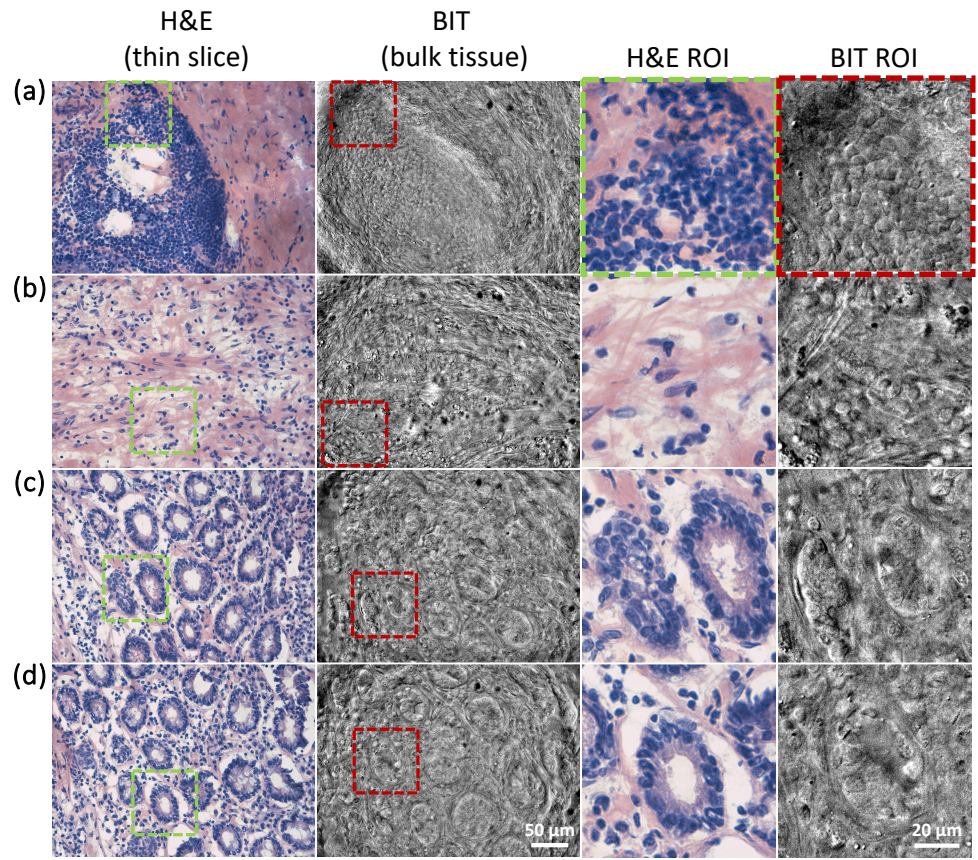

Fig. 5. (a)-(b) Pancreas and (c)-(d) duodenum imaged with conventional frozen section with H&E staining (left) and with BIT imaging of the adjacent bulk, unstained tissue face (right). Zoomed regions of interest indicated in green and red boxes are expanded to compare high-resolution H&E and BIT imaging.

## 4. Conclusion

Back-illumination interference tomography is a novel microscopy technique that utilizes a demagnified on-axis source imaged beyond the objective lens focal plane. Our results suggest that back-scattered light from this spatially-confined source image provides partially coherent, transmission-like illumination to objects at the focal plane. This condition leads to interferometric contrast, following the theory developed by Streibl [19]. Importantly, the interference contrast generated by BIT highlights axial phase gradients from small scattering objects. While similar to OBM in its goal of imaging bulk, unlabeled tissue, OBM utilizes off-axis, net oblique illumination to produce intensity changes encoded by lateral phase gradients. One can expect that the combination of BIT and OBM into a single microscope would improve the capture of highly detailed cellular images. The scattering properties of the medium are fundamental to the BIT mechanism, and the optimization of BIT contrast as a function of sample optical properties is an area that requires further investigation. Our initial studies indicate that BIT provides strong contrast in a wide variety of models up to 50 $\mu m$ deep. Imaging quality and optical sectioning as a function of depth is another important area to research, as the spatial coherence will degrade and background scattered light will increase for deep objects. Another area for future work is the development of quantitative phase imaging with BIT through the combination of images from different positions. Finally, artificial intelligence models could enable the rapid generation of 3D, virtually-stained H&E images of unprocessed bulk tissues. BIT provides optically-sectioned images of bulk tissue while circumventing the need for ultrafast lasers, beam scanning, and reference interference beam paths. The imaging speed is only limited by the image sensor frame rate. Given the simplicity of the technique and the high-contrast imaging capabilities demonstrated over a wide range of applications, BIT is a powerful tool for label free bulk tissue imaging.

## 5. Backmatter

**Funding.** Funding is provided through grants and gifts from the Gates Foundation and Fifth Generation Inc.

**Acknowledgments.** We want thank Alex S. Baras, MD, PhD for his generosity and help obtaining *ex vivo* human tissue specimens, and Timothy D. Weber, PhD for sharing his model for PSF simulations.

**Disclosures.** The authors are co-inventors on a provisional patent application assigned to Johns Hopkins University and Boston University. They may be entitled to future royalties from intellectual property related to the technologies described in this article.

**Data availability.** Data underlying the results presented in this paper are not publicly available at this time but may be obtained from the authors upon reasonable request.

**Supplemental document.** See Supplement 1 for supporting content.

## References

1. D. Huang, E. A. Swanson, C. P. Lin, *et al.*, "Optical Coherence Tomography," in *1Science,* vol. 254 (991), pp. 1178–1181.
2. A. Dubois, K. Grieve, G. Moneron, *et al.*, "Ultrahigh-resolution full-field optical coherence tomography," Appl. Opt. **43** (2004).
3. L. Wang, R. Fu, C. Xu, and M. Xu, "Methods and applications of full-filed optical coherence tomography: a review," J. Biomed. Opt. **27**, 1–26 (2022).
4. D. S. Gareau, "In Vivo confocal microscopy in turbid media," OHSU Digit. Commons: Scholar Arch. (2006).
5. P. J. Dwyer, C. A. DiMarzio, M. Rajadhyaksha, *et al.*, "Confocal reflectance theta line-scanning microscope for imaging human skin," Opt. Lett. **31**, 942–944 (2006).
6. M. Rajadhyaksha, R. R. Anderson, and R. H. Webb, "Video-rate confocal scanning laser microscope for imaging human tissues in vivo," Appl. Opt. **38**, 2105 (1999).
7. P. J. Dwyer, C. A. DiMarzio, J. M. Zavislan, *et al.*, "Confocal reflectance theta line scanning microscope for imaging human skin in vivo." Opt. Lett. **31**, 942–944 (2006).

8. I. Saknite, J. R. Patrinely, Z. Zhao, *et al.*, "Association of Leukocyte Adhesion and Rolling in Skin with Patient Outcomes after Hematopoietic Cell Transplantation Using Noninvasive Reflectance Confocal Videomicroscopy," JAMA Dermatol. **158**, 661–669 (2022).
9. F. Zernike, "Phase contrast, a new method for the microscopic observation of transparent objects," Physica **9**, 686–698 (1942).
10. F. Zernike, "How I discovered phase contrast," Science **121**, 345–349 (1955).
11. Y. K. Park, C. Depeursinge, and G. Popescu, "Quantitative phase imaging in biomedicine," Nat. Photonics **12**, 578–589 (2018).
12. M. E. Kandel, C. Hu, G. Naseri Kouzehgarani, *et al.*, "Epi-illumination gradient light interference microscopy for imaging opaque structures," Nat. Commun. **10**, 1–9 (2019).
13. T. H. Nguyen, M. E. Kandel, M. Rubessa, *et al.*, "Gradient light interference microscopy for 3D imaging of unlabeled specimens," Nat. Commun. **8** (2017).
14. T. N. Ford, K. K. Chu, and J. Mertz, "Phase-gradient microscopy in thick tissue with oblique back-illumination," Nat. Methods **9**, 1195–1197 (2012).
15. P. Ledwig and F. E. Robles, "Epi-mode tomographic quantitative phase imaging in thick scattering samples," Biomed. Opt. Express **10**, 3605–3621 (2019).
16. F. Robles, T. Abraham, P. Casteleiro Costa, *et al.*, "Label- and slide-free tissue histology using 3D epi-mode quantitative phase imaging and virtual H&E staining," Optica **10**, 12–15 (2023).
17. G. N. McKay, N. Mohan, and N. J. Durr, "Imaging human blood cells in vivo with oblique back-illumination capillaroscopy," Biomed. Opt. Express **11**, 2373–2382 (2020).
18. V. Mazlin, O. Thouvenin, S. Alhaddad, *et al.*, "Label free optical transmission tomography for biosystems: intracellular structures and dynamics," Biomed. Opt. Express **13**, 4190 (2022).
19. N. Streibl, "Three-dimensional imaging by a microscope," J. Opt. Soc. Am. A **2**, 121 (1985).
20. C. J. R. Sheppard and X. Q. Mao, "Three-dimensional imaging in a microscope," J. Opt. Soc. Am. A **6**, 1260 (1989).
21. T. D. Weber, "Transillumination Techniques in Ophthalmic I maging," Boston Univ. Coll. Eng. p. 89 (2013).
22. G. N. McKay, T. L. Bobrow, S. Kalyan, *et al.*, "Optimizing white blood cell contrast in graded-field capillaroscopy using capillary tissue phantoms," SPIE BiOS, 2020 p. 27 (2020).
23. G. N. McKay, L. Huang, T. L. Bobrow, *et al.*, "A model for generating paired complete blood count and oblique back-illumination capillaroscopy data in tissue-realistic microfluidic chambers," SPIE BiOS, 2022 p. 30 (2022).
24. G. N. McKay, L. H. Pecker, A. R. Moliterno, *et al.*, "A portable, dual-channel oblique back-illumination capillaroscope for in vivo human blood cell imaging in hematology clinics," SPIE BiOS, 2022 p. 3 (2022).